\documentclass[journal]{IEEEtran}

\usepackage{caption,subcaption,float}
\usepackage{graphicx}
\usepackage{multirow}
\usepackage{amsmath,fixmath,amsfonts}
\usepackage{array}
\usepackage{cite}
\begin{document}
\title{Latent Factor Modeling of Users Subjective Perception for Stereoscopic 3D Video Recommendation}

\author{Balasubramanyam~Appina$^{1}$,~\textit{Member},~\textit{IEEE},~Mansi~Sharma$^{2}$,~\textit{Member},~\textit{IEEE},~Santosh~Kumar$^{2}$. 
\thanks{$^1$The author is with Department of Electronics and Communication Engineering, Indian Institute of Information Technology, Design and Manufacturing Kancheepuram, Chennai, India, 600127.

E-mail: appina@iiitdm.ac.in.

$2$The authors are with Department of Electrical Engineering, Indian Institute of Technology Madras, Chennai, India, 600036.

E-mail: \{ee15b110@smail,~mansisharma@ee\}.iitm.ac.in.
}}

\maketitle

\begin{abstract}
Numerous stereoscopic 3D movies are released every
year to theaters and created large revenues. Despite the
improvement in stereo capturing and 3D video post-production
technology, stereoscopic artifacts which cause viewer discomfort
continue to appear even in high-budget films. Existing automatic
3D video quality measurement tools can detect distortions in
stereoscopic images or videos, but they fail to consider the
viewer’s subjective perception of those artifacts, and how these
distortions affect their choices. In this paper, we introduce a novel
recommendation system for stereoscopic 3D movies based on a
latent factor model that meticulously analyse the viewer’s subjective
ratings and influence of 3D video distortions on their preferences.
To the best of our knowledge, this is a first-of-its-kind model
that recommends 3D movies based on stereo-film-quality ratings
accounting correlation between the viewer’s visual discomfort
and stereoscopic-artifact perception. The proposed model
is trained and tested on benchmark Nama3ds1-cospad1 and
LFOVIAS3DPh2 S3D video quality assessment datasets. The
experiments revealed that resulting matrix-factorization based
recommendation system is able to generalize considerably better
for the viewer’s subjective ratings.
\end{abstract}

\begin{IEEEkeywords}
Stereoscopic 3D movies, recommender system, matrix factorization, latent factor models, subjective stereo quality assessment, visual discomfort
\end{IEEEkeywords}
\IEEEpeerreviewmaketitle

\section{Introduction}
\label{sec:intro}
The audience of 3D films and virtual reality content is
growing, as most of the films or YouTube videos have been
released in the stereoscopic 3D (S3D) format today. There
are three popular approaches to generate a stereoscopic 3D
video: 1) Scene acquisition using a stereo camera, 2) 2D-to-
3D video conversion, which means the creation of left and
right eye views from the original source video, 3) Rendering,
which is the process of synthesizing views by means of 3D
reconstruction or employing global 3D models and computer
vision techniques \cite{smolic2011three, sharma2014flexible,sharma2017uncalibrated,sharma2016novel,Cheng2012}. 

Despite advances in technology, there are numerous sources of visual artifacts to appear in the created stereoscopic picture/video \cite{antsiferova2017influence,website:Intro2}. A comprehensive study of visual artifacts in S3D content has been carried out at MSU Graphics \& Media Lab, Moscow State University, under VQMT3D project \cite{antsiferova2017influence,website:Intro2} in cooperation with IITP RAS. The research study identified potential artifacts in several popular Hollywood S3D movies. The artifacts like disparity, scale, color, sharpness mismatches or temporal asynchrony, cardboard, crosstalk effects are prominent in the S3D 3DTV content. Besides, different types of artifacts at various stages of the content delivery affect S3D video. The compression, blur and frame-freeze distortions influence 3D video in format-conversion and representation stage, and in the coding and transmission stage \cite{gotchev2011three}. Zeri and Livi \cite{zeri2015visual} interviewed 854 people.  They recognized frequent symptoms like eye strain, blurred vision and a burning sensation after watching 3D movies in theaters. Even high-budget films, like \textit{Pirates of the Caribbean, Dolphin Tale, The Three Musketeers, The Avengers, etc.}, contain scenes with geometric and color impairments, camera rotation difference, shift vertical variation between the left and right views. An important research study conducted by Barreda et al. \cite{barreda2014visual} on 3D content using psycho-physiological methods establish complex effects of visual discomfort over 3DTV viewer's emotional arousal, which leads to problems like headache, nausea, fatigue and eye strain, etc. The compression artifacts and their variation with a depth range of 3D displays noticeably affects viewer's perception \cite{appina2017subjective,jumisko2011subjective,sanchez2018performance,urvoy2012nama3ds1}. 

The most reliable way to reduce such distortions is to correct and enhance the stereoscopic-content quality during production. But the correction process is extremely labor-intensive and heavily relies on the degree of automation and on the workflow which is not cost efficient. The algorithms for the automatic detection of such artifacts and quality assessment are emerging \cite{bokov2014automatic,appina2019study}. However, measuring frequency and intensity of an artifact does not account how painful it can be for the viewer. Therefore, it is critical to consider subjective perception ratings for artifacts, that is, which egregious distortions affect a viewer notably and which distortions are within tolerable limits of his/her visual comfort. 

In this paper, we proposed a novel recommendation system for S3D movies. The well-controlled subjective experiments and careful statistical analysis conducted by the most studies establish that discomfort is greater for some specific distortions than for others when viewing stereo video \cite{li2015visual,website:Intro2}. Mainly the influence is from the content itself. We observed most significant information for designing a recommendation system for S3D movies is that describe the viewer's perceptual discomfort with the particular distortion types. Despite enough advances in image/video objective quality assessment techniques \cite{speranza2006effect,chen2013full,Flosim3D2017,appina2018}, it is difficult to propagate the same achievement for S3D video because automatic estimation of relevant characteristics for problems that cause visual discomfort is a nontrivial problem. We wonder when even very simple yet reliable metrics measure several problems affecting stereo quality on the fly. Thus, it is crucial to account subjective ratings for healthy and reasonable 3D video watching as well as properly designing of recommendation system.

Our recommendation system is built on the latent factor model that rely on viewer-movie ratings. Given a set of pristine and distorted S3D videos and their subjective ratings, our latent factor model that is based on matrix factorization map viewer's and 3D videos to 
a set of latent features. The problem of predicting perceptual quality rates of S3D video is formulated as a matrix completion problem for the user-movie rating matrix. Our system rate the S3D videos in accordance with the user's discomfort level. Our model recommendation mechanism can easily integrate within Netflix matrix factorization methods, which is the most important class of collaborative filtering approaches. The proposed recommendation system will be very useful in reducing the flood 
of low-quality 3D content online by ratings stereo 3D more-consistent with quality. The encouraging results obtained by statistical analysis of the proposed model conducted on benchmark Nama3ds1-cospad1 \cite{urvoy2012nama3ds1} and LFOVIAS3DPh2 \cite{appina2019study} S3D data demonstrate its potential for generating accurate predictions.

\section{Related Work} 
\label{RW}
A comprehensive survey of algorithms used by Netflix for its Recommendation System 
is found in a paper written by Leidy Esperanza MOLINA FERNÁNDEZ \cite{fernandez2018recommendation}. It covered Collaborative Filtering, Content-based Filtering, model-based SVD, PCA, and Probabilistic Matrix Factorization techniques. The paper explains a movie recommendation mechanism build within Netflix on the Matrix Factorization (MF) \cite{koren2009matrix} approach that learns the latent preferences of users and movies from the ratings and make a prediction of the missing ratings estimating the dot product of the latent factors \cite{fernandez2018recommendation,koren2009matrix}. 

Lu et al. \cite{lu2018convolutional} applied MF model for computing vector representations of words. Their work demonstrated how a convolutional neural network can be integrated into MF model to produce interpretable recommendations. Lee et al. \cite{lee2016content} developed a demo model that considers latest uploaded YouTube videos. Here the collaborative filtering approach is not much applicable since it relies on aggregate user behavior. Instead, they modeled recommendation problem as a video content-based similarity learning. They learned deep video embeddings and predict ground-truth video relationships from the trained model. However, this approach is built up purely based on video content signals.

YouTube provides a vast collection of 2D videos. In contrast since 2009, YouTube offers users an interesting feature to upload two-channel stereo videos for 3D viewing experience. YouTube flash players can support anaglyph videos in red/cyan, green/magenta or blue/yellow layout and follow row/column interlaced display on the screen. The 3D content on YouTube appears (or display) in accordance with the relevance order. Tsingalis et al. \cite{tsingalis2014statistical} presented a study on YouTube recommendation graphs of 2D and 3D videos. They studied the statistical relevance or recommendation properties of social network sites like Facebook, Tweeter and Flickr, such as power-law distribution. Also, they analyzed clustering methods to understand the existence of media content groups. Davidson et al. \cite{davidson2010youtube} discussed in details about the recommendation system in use at YouTube. The study reveals YouTube recommends personalized sets of videos to users based on their previous activity on the web. They discussed some unique challenges YouTube faces for video endorsement and how to address them. Covington et al. \cite{covington2016deep} describes a YouTube system at a high level and center their study about substantial performance improvements brought by deep learning. They presented deep architecture built on deep candidate generations and a separate ranking model for recommending YouTube videos.

Estrada and Simeone \cite{estrada2017recommender} developed a recommender system for guiding physical object substitution in virtual reality. This user-perception based recommender approach allows them to watch the physical world whilst navigating the virtual environment through a video feed. The user identifies the location of object placement in the surroundings given the feedback.

Niu et al. \cite{niu2017temporal} presented a video recommendation system based on the affective analysis of the users. Their subjective model evaluates feature of emotion fluctuation based on the Grey Relational Analysis (GRA). Certain video features are extracted and mapped to the well-known Lovheim emotion-space specifying prominent human feelings, patterns, attitudes and behaviour such as Anger, Distress, Surprise, Fear, Enjoyment, Shame, Interest, and Contempt. GRA-based recommendation method is developed under the Fisher model to analyse extracted emotions as factors.

Zhang et al. \cite{zhang2013improving} developed a recommendation system for Mobile AR application incorporating user's preferences, location and temporal information in an aggregated random walk algorithm. Their system predicts user's preferences modifying the graph edge weight and computing the rank score. Similarly, Shi et al. \cite{shi2015novel} predicts individual location recommendation, Chatzopoulos and Hui \cite{chatzopoulos2016readme} anticipates object recommendation in Mobile AR environments.

\begin{figure*}[t] 
\centering
\begin{subfigure}[b]{0.45\textwidth}
\includegraphics[height=6cm]{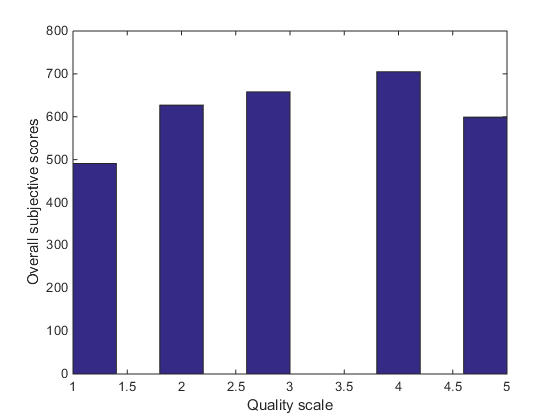}
\subcaption{Nama3ds1-cospad1 dataset \cite{urvoy2012nama3ds1}.\\ $\mu$ = 3.09, $m$ = 3, $\sigma$ = 1.35.}
\label{fig:IRCCYN_Dist}
\end{subfigure}
\begin{subfigure}[b]{0.45\textwidth}
\includegraphics[height=6cm]{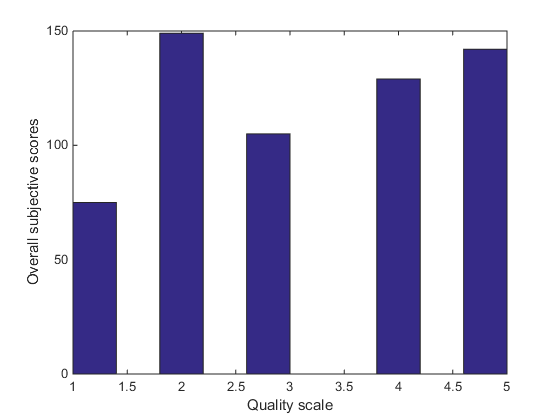}
\subcaption{LFOVIAS3DPh2 dataset \cite{appina2019study}.\\ $\mu$ = 3.19, $m$ = 3, $\sigma$ = 1.37.}
\label{fig:LFOVIAPh2S3D}
\end{subfigure}
\caption{\small Subjective score distribution of dataset. The $\mu$, $m$ and $\sigma$ denote the statistical measures (mean ($\mu$), median ($m$), standard deviation ($\sigma$)) of the subjective ratings.}
\label{fig:subjectivedist}
\end{figure*}

\section{Mathematical Modeling of S3D videos Recommendation System} 
\label{MM}
We proposed a novel recommendation system for stereoscopic 3D videos based on a MF model. In the proposed model, viewer's and S3D movies are mapped to a joint latent factor space. The row or column associated to a specific viewer or S3D movie is referred as the latent factors. In the mapped latent factor space of dimensionality, say $f$, the viewer-movie ratings are analyzed as inner products. Suppose each S3D movie $i$ is associated with a latent vector $q^{m}_i \in R^{f}$, and each user $u$ is associated with a latent vector $p^{u}_j \in R^{f}$. In the proposed problem formulation, for a given movie $i$, the elements of $q^{m}_i$ estimate the extent to which the S3D movie holds those factors, whether distorted with a particular artifact or free from that. For a given user $u$, the elements of $p^{u}_j$ determine the extent of user acceptance in S3D movies that are high on the corresponding factors, again, whether distorted with a particular artifact or not. The model approximates viewer $u's$ rating of S3D movie $i$ by measuring resulting dot product, $\hat{r}_{ui}=q^{m^{T}}_i p^{u}_j$. The dot product captures interconnection between the viewer $u$ and S3D movie $i$, that is, the viewer's overall acceptance/tolerance in the particular distortion viewing the movies. Once the mapping is computed for each S3D movie and viewer to factor vectors $q^{m}_i, p^{u}_j \in R^{f}$, the proposed model easily determines the rating a viewer will give to any S3D movie with distortions by using $\hat{r}_{ui}$.

We avoided imputation in proposed model \cite{bertsimas2017predictive}. The observed ratings are modeled directly as suggested by \cite{koren2009matrix,paterek2007improving} and avoided overfitting through the regularization. On the set of known matrix ratings, the regularized squared error is minimized
to learn the factor vectors $q^{m}_i,p^{u}_j$ as

\begin{equation}
\underset{\hat{p},\hat{q}}{min} \sum_{(u,i) \in \mathfrak{S}}(r_{ui}-q^{m^{T}}_i p^{u}_j)^{2}+ \lambda (||q^{m}_i ||^{2} + || p^{u}_j||^{2} )    
\label{Eq2}
\end{equation}

where, $\mathfrak{S}$ is the training set of $(u,i)$ pairs for which $r_{ui}$ is known.

To make matrix factorization approach more effective in our proposed application-specific requirements, we add biases in capturing the full ratings of the observed signals
\begin{equation}
\hat{r}_{ui}=\mu + b_i + b_{u} + q^{m^{T}}_i p^{u}_j
\label{Eq5} 
\end{equation}
The observed rating in (\ref{Eq5}) is broken down into its four components: global average (or mean), 3D movie bias, viewer bias, and viewer-movie interaction. This allows each component to represent only the part of an observed signal relevant to it. 
The model is learned by minimizing the squared error function as
\begin{multline}
\underset{\hat{p},\hat{q},\hat{b}}{min} \sum_{(u,i) \in \mathfrak{S}}(r_{ui}-\mu - b_i - b_{u}-q^{m^{T}}_i p^{u}_j)^{2}+ \lambda \\ (||q^{m}_i ||^{2} + || p^{u}_j||^{2} + b^{2}_i + b^{2}_u)    
\label{Eq6}
\end{multline}

The stochastic gradient descent algorithm \cite{koren2009matrix,koren2008factorization,jin2018regularizing} is used to optimize equation (\ref{Eq6}). For better accuracy in prediction, the algorithm loops through all ratings in the training data and estimate the model parameters. The system estimates $\hat{r}_{ui}$ for each given training case. The prediction error is determined as 
\begin{equation} 
E_{ui} = {r}_{ui} - \mu - b_i - b_{u} - q^{m^{T}}_i p^{u}_j
\end{equation} 
The parameters are updated as 
\begin{equation} 
b_i \leftarrow  b_i + \varsigma  (E_{ui} - \lambda b_i)
\end{equation} 
\begin{equation} 
b_u \leftarrow  b_u + \varsigma  (E_{ui} - \lambda b_u)
\end{equation} 
\begin{equation} 
q^{m}_i \leftarrow  q^{m}_i + \rho (E_{ui} p^{u}_j - \lambda q^{m}_i)
\end{equation} 
\begin{equation} 
p^{u}_j \leftarrow  p^{u}_j + \rho (E_{ui} q^{m}_i - \lambda p^{u}_j)
\end{equation} 
where, $\rho$ and $\varsigma$ specify constant magnitudes that accounts proportion by which parameters are modified in the opposite direction of the gradient.

The objective of our matrix factorization model is to the predicts the unknown future S3D video ratings, from the learned model obtained by fitting the earlier observed ratings. We determined the regularization constant $\lambda$ by cross-validation \cite{mnih2008probabilistic}.

\section{Results and Discussion}
\begin{figure*}[t] 
\centering
\begin{subfigure}[b]{0.4\textwidth}
\includegraphics[width=6.5cm,height=4cm]{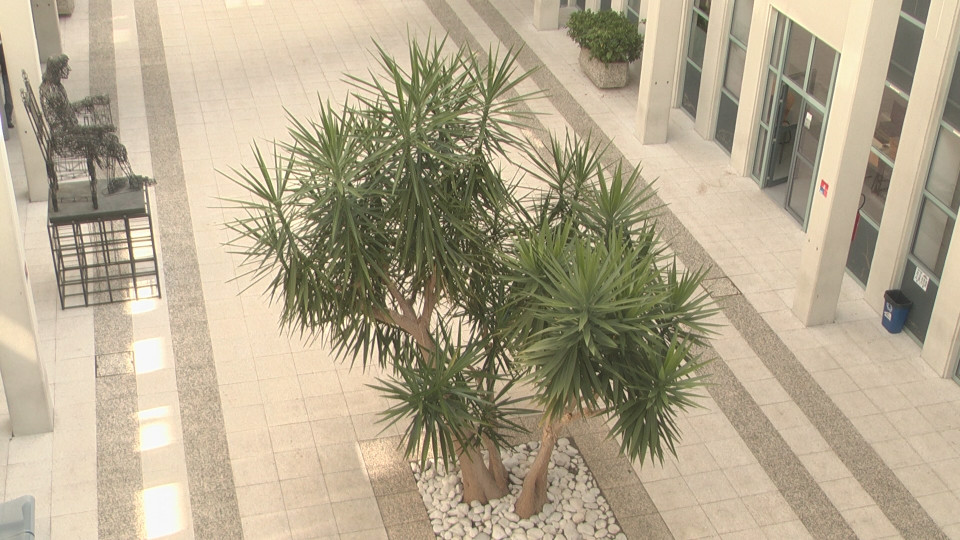}
\subcaption{Pristine stereoscopic video frame.}
\label{fig:PristineFrame}
\end{subfigure}
\begin{subfigure}[b]{0.4\textwidth}
\includegraphics[width=6.5cm,height=4cm]{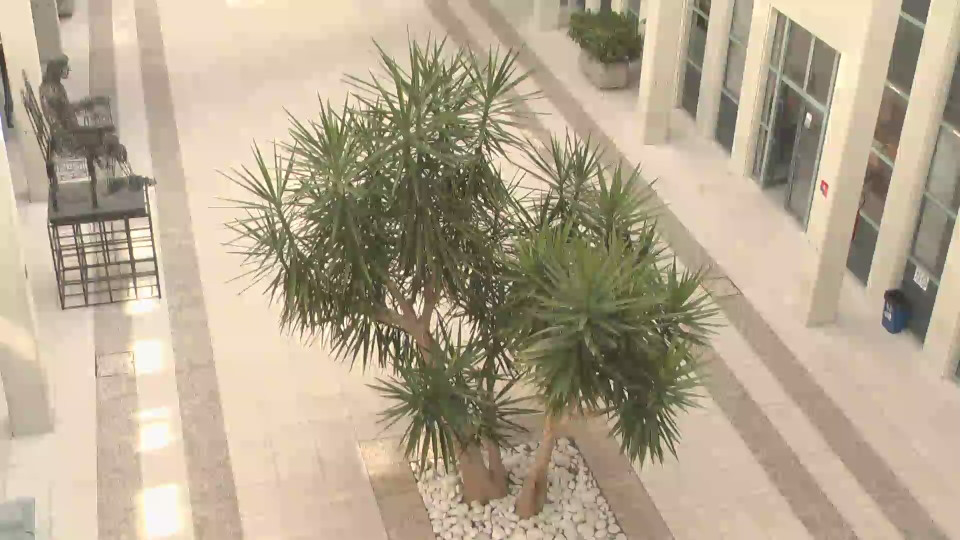}
\subcaption{Distorted stereoscopic video frame.}
\label{fig:DistortedFrame}
\end{subfigure}
\\
\begin{subfigure}[b]{0.44\textwidth}
\includegraphics[width=7.5cm,height=6cm]{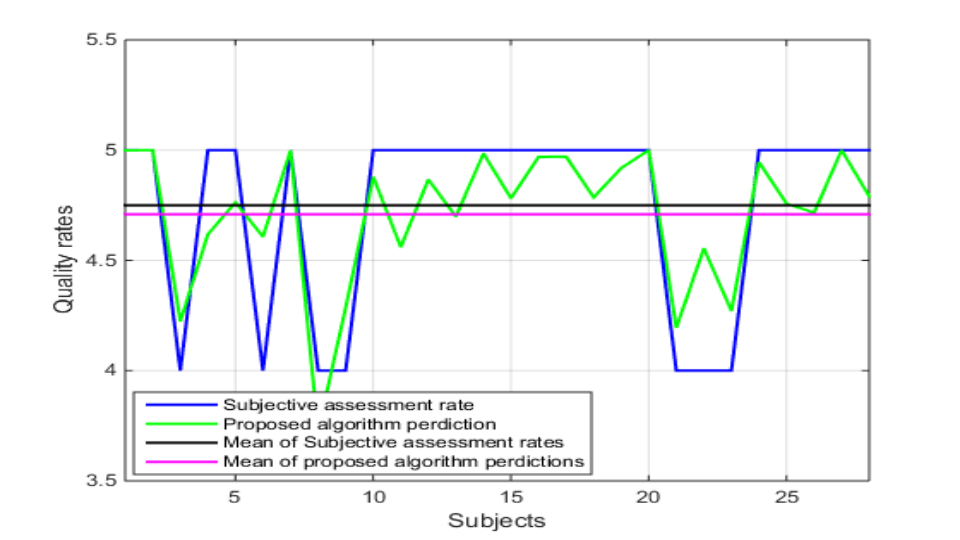}
\subcaption{Pristine stereoscopic video.}
\label{fig:subjectiveandobjectivePristine}
\end{subfigure}
\begin{subfigure}[b]{0.45\textwidth}
\includegraphics[width=7.5cm,height=6cm]{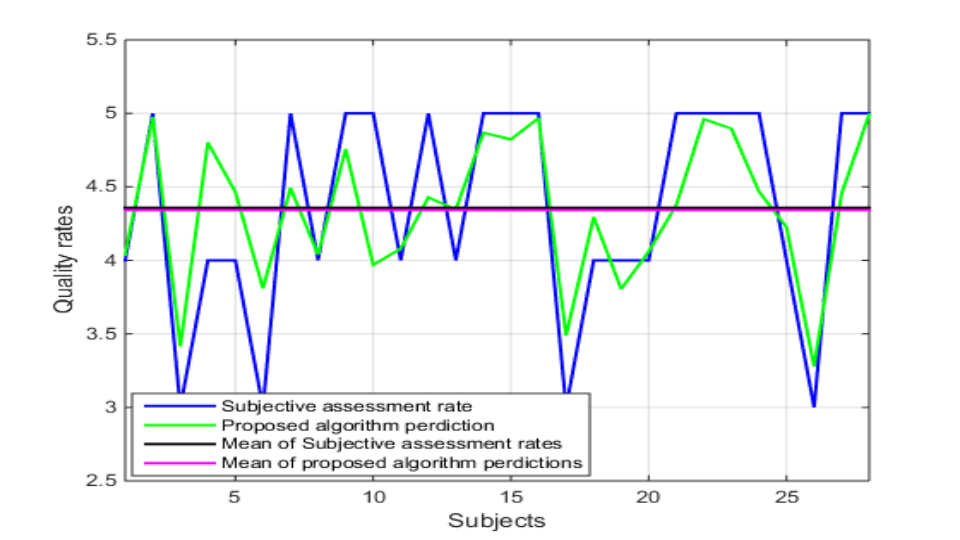}
\subcaption{Distorted stereoscopic video.}
\label{fig:subjectiveandobjectiveDistorted}
\end{subfigure}
\caption{\small Perceptual opinion score of each subject and proposed algorithm prediction on pristine and distorted `Hall' S3D videos from the Nama3ds1-cospad1 \cite{urvoy2012nama3ds1} dataset.}
\label{fig:subjectiveandobjective}
\end{figure*}
The efficacy of the proposed algorithm is evaluated on the Nama3ds1-cospad1 \cite{urvoy2012nama3ds1} and LFOVIAS3DPh2 \cite{appina2019study} S3D video datasets. Nama3ds1-cospad1 database has 10 reference and 100 test S3D video sequences. The video sequences have a resolution of $1920 \times 1080$ and saved in .avi container. The frame rate is 25 fps and a duration of either 16 sec or 13 sec. The database is a combination of H.264 and JP2K, scaling and down sampling distorted S3D video sequences. These artifacts are applied symmetrically on each view of an S3D video and published the Difference Mean Opinion Score (DMOS) scores as subjective scores. Human assessment on perceptual quality was performed in single stimulus continuous quality evaluation (SSCQE) with hidden reference method. They have used 5 scales to rate the perceptual quality of an S3D video and 28 subjects involved in the study. They have published each subject quality score and an overall mean quality score of the dataset. The LFOVIAS3DPh2 S3D video dataset has 12 pristine sequences with good variety of structure, texture, depth and temporal information. The video sequences have a resolution of $1920 \times 1080$ and duration of 10 seconds with a frame speed of 25 fps. They created 288 test stimuli by introducing the H.264 and H.265 compression, blur and frame freeze distortions. The dataset is a combination of symmetric and asymmetric S3D videos. They have used SSCQE method to perform the subjective study and 20 subjects involved in the study. They published each subject perceptual quality score and final DMOS of the dataset. Figure \ref{fig:subjectivedist} shows the viewer perceptual score distribution of Nama3ds1-cospad1 and LFOVIAS3DPh2 S3D video datasets. From the plot, it is clear that both datasets are diverse in perceptual video quality range. Also, the estimated statistics ($\mu$, $m$ and $\sigma$) of each dataset are consistent and followed the observed trend in the perceptual quality of S3D videos.

Figure \ref{fig:PristineFrame} shows the $1^{st}$ frame from left view of the pristine `Hall' S3D video from Nama3ds1-cospad1 dataset. Figure \ref{fig:DistortedFrame} shows the $1^{st}$ frame of H.264 (Quantization Parameter = 38) compressed S3D video of the corresponding reference view. Figures \ref{fig:subjectiveandobjectivePristine} and \ref{fig:subjectiveandobjectiveDistorted} show the distribution of subjective assessment rates and proposed algorithm predicted perceptual quality rates of a pristine and distorted S3D video, respectively. From the plot it is clear that the proposed algorithm accurately predicts the subjective quality rates of pristine and distorted videos. Also, the deviation between average scores of subjective rates and the proposed algorithm predictions is very less. The plot clearly demonstrates the proposed algorithm efficacy to model the perceptual subjective quality ratings of a given S3D video.

\begin{table*}[!htbp]
\small
\caption{\small Performance evaluation of proposed algorithm on Nama3ds1-cospad1 \cite{urvoy2012nama3ds1} and LFOVIAS3DPh2 \cite{appina2019study} video dataset subjective scores.}
\centering 
\begin{tabular}{|c| c| c|c| c| c|c| } 
\hline
  \multirow{2}{*}{\bf Score} & \multicolumn{3}{c|}{\bf Training Set}& \multicolumn{3}{c|}{\bf Testing Set}\\
 \cline{2-7} 
  &  {\bf LCC}  & {\bf SROCC} & {\bf RMSE}&  {\bf LCC}  & {\bf SROCC} & {\bf RMSE}\\
\hline
Nama3ds1-cospad1 \cite{urvoy2012nama3ds1} & 0.8873&0.8858&0.6903&0.8753&0.8700&0.7527 \\
\hline
LFOVIAS3DPh2 \cite{appina2019study} &0.8966 &0.8911 &0.5809 &0.8585 &0.8288 &0.6522 \\
\hline
\end{tabular}
\label{tab:Results}
\end{table*}

\begin{table}[h]
\caption{\small Proposed algorithm performance on each distortion type of LFOVIAS3DPh2 S3D video dataset \cite{appina2019study}.}
\centering 
\begin{tabular}{|c| c| c| c|c|c|} 
\hline
  Type&  H.264 & H.265& Blur&FF&Overall \\
\hline
LCC & 0.910 & 0.904 & 0.877&0.908 &0.858 \\
\hline
 SROCC & 0.902& 0.914& 0.828& 0.899& 0.828\\
\hline 
RMSE  & 0.357& 0.384& 0.423&0.371 & 0.652\\
\hline 
\end{tabular}
\label{tab:EachDistType}
\end{table}

The performance of the proposed algorithm is measured using the Linear Correlation Coefficient (LCC), Spearman Rank Order Correlation Coefficient (SROCC) and Root Mean Square Error (RMSE). LCC indicates the linear dependence between two quantities. The SROCC measures monotonic relationship between two input sets. RMSE measures the magnitude error between estimated scores and subjective scores. Higher LCC and SROCC values indicate good agreement between subjective and objective measures, and lower RMSE signifies more accurate prediction performance. For both the databases, 80\% of the human opinion scores is used for proposed algorithm training and the remaining samples are used for testing. In other words, the non-overlapped training and test sets are obtained by partitioning the set of available human opinion scores in the 80:20 proportion. We performed the random assignment for 100 trials of each epoch for statistical consistency and repeated it for 200 epochs. Finally, we calculated the mean of the  LCC, SROCC and RMSE measures of all epochs to report the performance analysis. Table \ref{tab:Results} shows the performance evaluation of the proposed algorithm on the training and test sets of Nama3ds1-cospad1 and LFOVIAS3DPh2 S3D video datasets. It is clear that the proposed algorithm shows robust performance across both the datasets. 
\begin{figure}[h]
  \includegraphics[width=\linewidth]{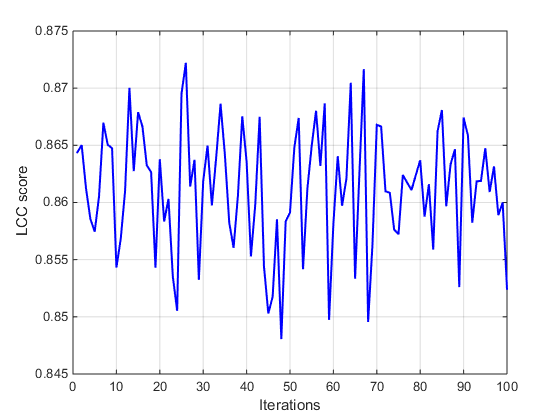}
  \caption{Variation of proposed algorithm LCC score over 100 trails of an epoch. Standard deviation of LCC score over 100 trails is $2 \times 10^{-4}$.}
  \label{fig:100It}
\end{figure}

\begin{figure}[h]
  \includegraphics[width=\linewidth]{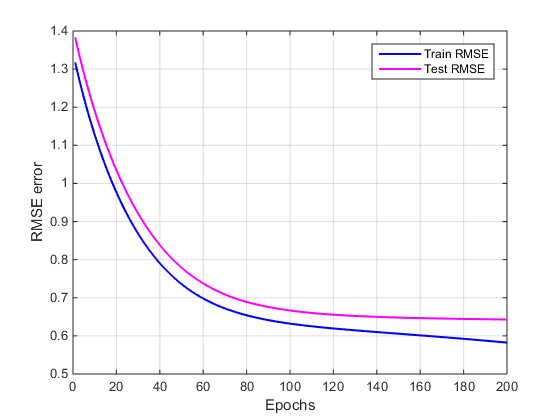}
  \caption{RMSE error variation across epochs.}
  \label{fig:RMSEscorevariation}
\end{figure}

Figure \ref{fig:100It} shows the LCC score variation of 100 iterations of an epoch. From the plot it is clear that the scores are consistent across all iterations, and further, we experienced the lower standard deviation $(2 \times 10^ {-4} $) of 100 LCC scores. Figure \ref{fig:RMSEscorevariation} shows the average training and test RMSE measure variation over 200 epochs. From the plot, it is clear that both the RMSE errors reduced with the number of epochs. These plots clearly demonstrate the proposed algorithm efficacy to estimate the human assessment quality of a given video. Table \ref{tab:EachDistType} shows the proposed method performance on each distortion type of LFOVIAS3DPh2 S3D video dataset. The proposed method clearly demonstrates the consistent performance across all distortion types of the LFOVIAS3DPh2 S3D video dataset. 


\section{Conclusion} 
\label{sec:con} 
This paper presented a novel recommendation system for S3D movies. This is a first attempt that accounts 3DTV viewer's subjective ratings for visual artifacts and analyse their degree of visual discomfort to predict ``rating'' or ``preference'' that the viewer's would give to the S3D movie. In this study, we considered common spatial and temporal distortion types; JPEG, Resolution reduction, Downsampling \& sharpening, Image sharpening, Blur, Frame-freeze, H.264 and H.265 compressions; that adversely affect S3D video signal at different stages of the content generation and delivery chain. Experimental results on 3DTV viewer's subjective study and parameter evaluation of latent factors demonstrate that the proposed matrix factorization based model improve accuracy of S3D video affective analysis and performance of recommendation. This model will be very useful for media-service providers like Netflix, Amazon, TiVo to recommend quality 3D videos and minimize flood of low-quality content based on the viewer's subjective perception, depending on their age groups and preferences.

We will further extend this recommendation system by considering the detail analysis of commercial S3D movies. The model will be improved by offering per-frame analysis of artifacts causing potential visual discomfort while viewing stereo films like large horizontal disparity, vertical parallax, crosstalk noticeability, cardboard effect, stuck-to-background objects, stereo window violation, depth continuity, etc. Such artifacts earn poor ratings according to the existing metrics. Combining objective and subjective scores will help reduce the error rate further while recommending new
stereo movies. 

Besides, we will perform affective analysis on the emotional reactions of 3DTV viewers while watching stereo 3D movies or virtual reality S3D content. We will account both subjective scores and brain-activity measurements to understand the dependencies between the degree of viewer discomfort and the intensity of the distortions. This will help to better classify viewer's from different age groups by their susceptibility to artifacts and movies content types. How this affect viewer's accumulation of discomfort caused by stereoscopic movies and influence recommendation ratings is an interesting endeavour of future study ?. In future work, we will account the percentage of viewers susceptible
to various distortions. We will design new experiments and work on evaluation models like probabilistic matrix factorization (PMF) to improve the predictive accuracy. We will experiment on the linear combination of predictions of multiple PMF models with
predictions of Restricted Boltzmann Machine (RBM) models.
This could significantly improve the accuracy of the blended solution.

\bibliographystyle{ieeetr}
\footnotesize
\bibliography{root.bib}
\end{document}